\documentclass[%
 reprint,
superscriptaddress,
 amsmath,amssymb,
 aps,
twocolumn]{revtex4-2}

\usepackage{graphicx}
\usepackage{dcolumn}
\usepackage{bm}
\usepackage[dvipsnames]{xcolor}
\usepackage{stackrel}
\usepackage{siunitx}
\usepackage{physics} 

\usepackage{amsfonts}

\usepackage{bbold}

\newcommand{\sophiem}[1]{}
\newcommand{\sophieh}[1]{}
\newcommand{\carlos}[1]{}
\newcommand{\adam}[1]{}

\newcommand{\brak}[1]{\langle #1 \rangle}

\newcommand{\hatrho}{\hat{\rho}}

\usepackage{hyperref}
\usepackage[shortlabels]{enumitem}

\usepackage{multirow}

\usepackage{lineno}

\graphicspath{{Figures/}}

\AtBeginDocument{%
  \setlength{\linenumbersep}{\dimexpr\oddsidemargin+0.48in-\linenumberwidth\relax}%
}

\begin{document}

\preprint{APS/123-QED}

\title{The ``Intensity'' Countoscope: \\ Measuring particle dynamics in real space from microscopy images}

\author{Sophie Hermann}%
\affiliation{%
CNRS, Sorbonne Université, Physicochimie des Electrolytes et Nanosystèmes Interfaciaux, F-75005 Paris, France
}%

\author{Seyed Saman Banarooei}
\affiliation{Department of Chemical and Biomolecular Engineering, Vanderbilt University, Nashville, TN, USA.}

\author{Adam Carter}
\affiliation{%
CNRS, Sorbonne Université, Physicochimie des Electrolytes et Nanosystèmes Interfaciaux, F-75005 Paris, France
}%

\author{Carlos A. Silvera Batista}
\affiliation{Department of Chemical and Biomolecular Engineering, Vanderbilt University, Nashville, TN, USA.}
\affiliation{Vanderbilt Institute for Nanoscale Science and Engineering, Vanderbilt University, Nashville, TN, USA.}

\author{Sophie Marbach}%
\affiliation{%
CNRS, Sorbonne Université, Physicochimie des Electrolytes et Nanosystèmes Interfaciaux, F-75005 Paris, France
}%
\email{sophie.marbach@cnrs.fr}

\begin{abstract}
Advances in intensity-based microscopy techniques have improved our ability to quantify particle motion at microscopic scales, enabling insight into diffusion and collective dynamics. Building on this foundation, we introduce a novel real-space approach that analyses intensity fluctuations within virtual observation boxes of variable size on microscopy images. By correlating these signals, we uncover distinct temporal regimes in the mean square changes of intensity, $\langle \Delta I^2(t) \rangle$, 
which are strongly dependent on the box size compared to the particle width. For small boxes or short timescales, $\langle \Delta I^2(t) \rangle$ scales with the mean-square displacement, while for longer timescales and  larger boxes, it scales with its square root. We develop a general theoretical framework that captures these regimes and explicitly apply it to a dilute colloidal suspension imaged with confocal microscopy as an experimental model system. This allows for a robust extraction of diffusion coefficients and physical insights into particle dynamics. Our method complements intensity-based and real-space analysis, offering a tool for studying individual and potentially collective behaviour directly from image intensities, even in systems where individual particles cannot be resolved.
\end{abstract}

\maketitle

Driven by the desire to understand particle motion, scientists rely on a zoology of methods to quantify motion at microscopic scales. To observe micrometric particles, the state-of-the-art is to acquire microscopy videos, imaging thousands of colloids or cells with single particle resolution. Then, data analysis can be sorted into three loose categories, depending on whether it is based on \textit{(i)} the intensity of light scattered by the particles on images~\cite{magde1972thermodynamic,berne1976}, \textit{(ii)} segmented features, such as particle positions and orientations~\cite{mackay2024}, or \textit{(iii)} reconstructed particle trajectories, found by linking particle positions from one frame to the next~\cite{crocker1996methods}. 
As an example, and as proposed by the seminal works of Einstein and Perrin~\cite{einstein1905molekularkinetischen,perrin2014atomes}, the diffusion coefficient, $D$, can then be inferred by computing a particle's mean-square displacement from the trajectories, since $\brak{\Delta x^2(t)} = \langle (x(t)-x(0))^2 \rangle = 2D t$, where $x(t)$ is the particle's position along one cartesian axis. Working at the level of intensities necessarily requires the least image processing, and is thus appealing. 

The history of quantifying diffusion coefficients from particle intensities merits discussion in the light of modern tools. 
In the years 1914-1916, Smoluchowski derived an important formalism~\cite{smoluchowski1914studien,smoluchowski1915uber,smoluchowski1916studien}, later verified by experiments~\cite{westgren1916bestimmungen,westgren1918koagulation}, to measure diffusion coefficients from correlation functions of number fluctuations. Indeed, the number of particles appearing within the finite field of view of a microscope fluctuates due to particles diffusing in and out of the field. Possibly due to technical difficulties in analyzing images at the time, this principle remained dormant for half a century. In the 1970s, thanks to technical developments in optics, two methods emerge based on this idea: fluorescence correlation spectroscopy (FCS)~\cite{magde1972thermodynamic,elson1974fluorescence} and dynamic light scattering (DLS)~\cite{berne1976}. Both methods directly correlate the intensity scattered by the few particles in the illuminated region to obtain features such as diffusion coefficient. FCS is remarkable as it is suited to the investigation of small molecules, and also reaction timescales~\cite{elson2011fluorescence}. DLS can further analyze motion at different spatial scales in Fourier space (k-space)~\cite{berne1976}. 
However, both these techniques do not work yet at the level of images. \adam{the difference between FCS and DLS is not clear here. (I think it's the illumination type and volume maybe?)}

In the early 2000s, Intensity correlation spectroscopy (ICS) was developed as the microscopy space equivalent of FCS/DLS~\cite{wiseman2015image,jameson2009fluorescence}. The general idea is to correlate pixel intensities to reconstruct diffusion coefficients~\cite{kolin2007advances}. Pixels can be correlated in space and time for different uses~\cite{wiseman2015image}, and maps of diffusion coefficients can be obtained in that way, which are especially useful to investigate e.g. cellular environments~\cite{scipioni2018local}. Most of these methods were developed for very small molecules in biological contexts~\cite{kolin2007advances}. Around the same time, ICS was developed in spatial k-space ICS~\cite{kolin2006k}, a premise for the establishment of dynamic differential microscopy (DDM)~\cite{cerbino2008differential}. In DDM, the Fourier transform of differences in images at different times is correlated, and allows one to obtain estimates of diffusion coefficients at multiple scales and further motion properties, with numerically efficient tools~\cite{giavazzi2014digital,lattuada2025hitchhiker}. All these techniques are especially useful at the nanoscales where segmentation of individual particles is not possible. 
\sophiem{read history on hitchhiker, add a DLS reference from Montpellier}

Still, there remain more ideas to investigate the full depth of Smoluchowski's vision. In fact, optical techniques have not ceased to improve our ability to image individual particles at smaller and smaller scales, with super-resolution and interferometric scattering microscopy now resolving individual molecular dynamics~\cite{comtet2021anomalous,ronceray2022liquid,hell1994breaking,rust2006sub,schmidt2021minflux}. 
This means that microscopy methods are poised for real-space analysis of the type of Smoluchowski. Recently, some of us have proposed to exploit image quality by analyzing particle number fluctuations in \textit{virtual} observation boxes~\cite{mackay2024}, instead of only the full field of view. Investigating boxes of varying sizes allows one to quantify motion at multiple scales~\cite{carter2025measuring}, and real-space analysis allows one to easily connect to physical mechanisms~\cite{mackay2025collective}. This highlights a gap in contemporary analysis techniques based on intensities: in principle, there should exist a method based on analyzing intensity signals in \textit{virtual} boxes of arbitrary size on an image. The versatility of DDM which operates on multiple scales in Fourier space highlights the potential of learning from multiple box sizes in real-space. 

Here we investigate the correlation functions of intensity signals $I(t)$ in \textit{virtual} observation boxes of varying sizes on microscopy images. We image dilute colloidal suspensions with confocal microscopy as a model system. The intensity fluctuations $\langle \Delta I^2(t) \rangle$ exhibit sharp signatures at short and long timescales, which differ according to the size ratio of the box and the point spread signal of a particle on an image. Briefly, for boxes smaller than the particle size, or at short time scales, the intensity correlations $\langle \Delta I^2(t) \rangle \sim  \brak{\Delta x^2(t)}$ while for boxes larger than the particle size and at longer time scales, $\langle \Delta I^2(t) \rangle \sim \sqrt{\brak{\Delta x^2(t)}}$. We derive a general theory, applicable to diverse optical systems, which collapses into simple expressions and completely captures experimental signals. This allows us on the one hand to extract experimental parameters such as the diffusion coefficient $D$ from $\langle \Delta I^2(t) \rangle$, and on the other hand, to learn physical behaviour. In particular, the different regimes in time can be explained in terms of diffusive particle exits/entrances in the box. Finally, we show how such analysis can robustly be conducted on systems even where we can not identify individual particles on images. Our method opens a door to further quantify individual and collective behaviour directly from image intensities in various settings.

\begin{figure*}[htbp]
    \centering
    \includegraphics[width=1.\textwidth]{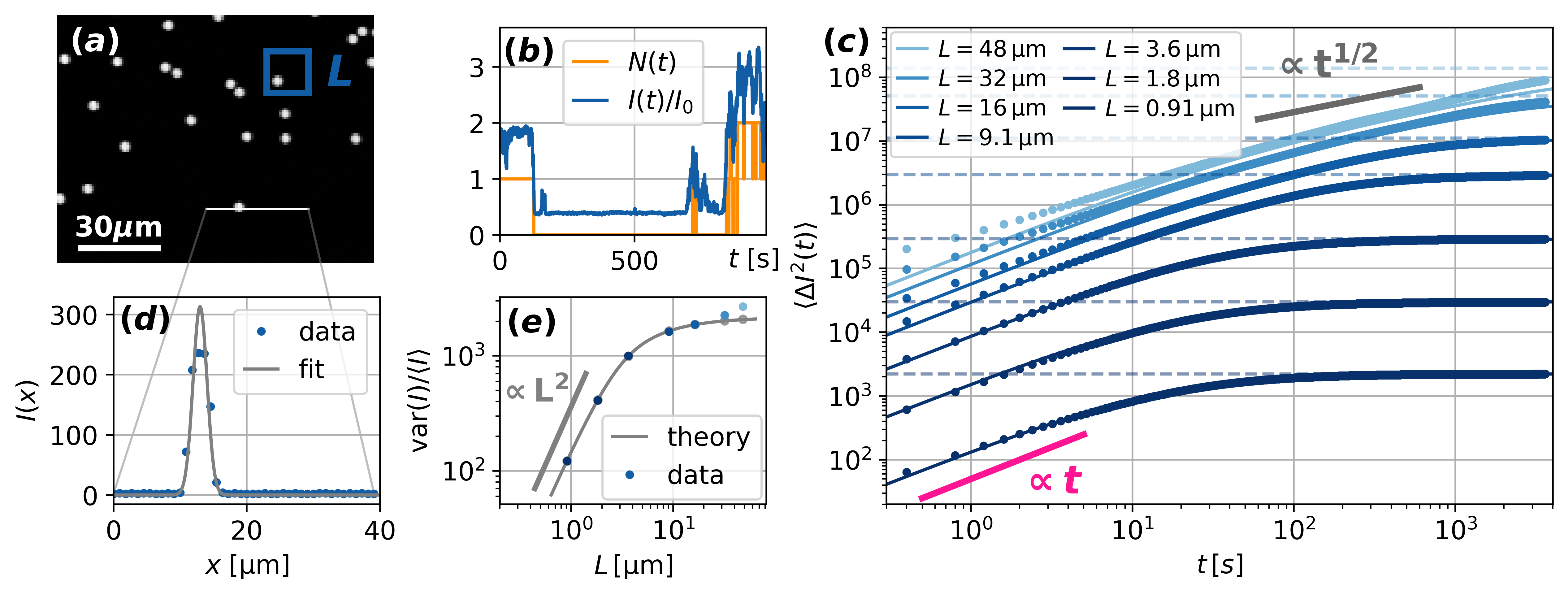}
    \caption{\textbf{The intensity countoscope to probe dynamics.} \textbf{(a)} Part of a microscopy image of quasi-2D diffusing fluorescent particles, overlayed with an observation box of size $L = 16~\unit{\micro \metre}$. \textbf{(b)} Temporal fluctuations of the particle number (orange) and total intensity (blue), rescaled by the average particle intensity $I_0$, in the box in (a). \textbf{(c)} Mean-squared change in intensity for experiments (points), theory (solid lines, Eq. \eqref{eq:I_Gaussian_Square}), and in the long time limit (dashed lines, Eq. \eqref{eq:msi_infty_theo}). Colours, from dark to light, indicate different box sizes $L$ ranging from $0.91~\unit{\micro \metre}$ (corresponding to one pixel) to $48~\unit{\micro \metre}$. \textbf{(d)} Intensity profile of a particle cross-section: experimental values (dots) and Gaussian fit (line), $I_0 \exp(-x^2/2\sigma^2)/(2\pi\sigma^2)$, where $\sigma$ and $I_0$, are determined from the fit in (e). \textbf{(e)} Intensity variance rescaled with mean intensity dependent on the observation box size for experiments (dots) and theory (line). Blue colours for the dots are chosen according to the box size as in (c).}
    \label{fig:1}
\end{figure*}

\section*{Intensity fluctuations in boxes in experimental data exhibit well defined varied features.}
We study the temporal intensity fluctuations in microscopy images. Our experimental model system consists of colloidal polystyrene (PS) particles, of $4~\unit{\micro\metre}$ diameter, which are gravitationally confined to the base of a fluidic cell
and form an effective 2D system (see also Methods, Sec. A). The particles are imaged with a fluorescent confocal microscope, see Fig.~\ref{fig:1}(a). We characterize intensity fluctuations in virtual observation boxes of size $L \times L$ by summing all pixel intensities in the corresponding area. Note that boxes are aligned with the pixel edges in the microscopy image and we consider only box sizes $L$ that are multiples of a pixel size.  In Figure \ref{fig:1}(b) this intensity $I(t)$  is shown for the exemplary box shown in Fig. \ref{fig:1}(a). For consistency checks, we also track individual particle centers using standard particle tracking~\cite{crocker1996methods} with the \textit{Trackpy} library \cite{trackpy}. This allows us to obtain individual particle trajectories, and also particle numbers $N(t)$ in the observation boxes. 

The intensity in a box $I(t)$ fluctuates in time as particles move near the edges of the box, or when a particle enters or exits the box. The temporal evolution of the intensity therefore closely follows the detected number of particles $N(t)$ in the observation box. However, the intensity profile has additional features: it takes continuous instead of discrete values it time; and has a global offset due to the background intensity. 

For each observation box, we calculate the mean square change of the intensity, given as
\begin{align}
\left< \Delta I(t)^2\right> &=\left< \big(I(t) - I(0)\big)^2\right>  \nonumber \\
&= 2 \left<  I^2\right>  - 2 \left< I(t) I(0)\right>, \label{eq:msi}  
\end{align}
where the average $\left< \cdot \right>$ denotes an averaging over all boxes of size $L$ and all time origins within the acquisition (see also Methods, Sec. B). 
Investigating intensity differences as $I(t) - I(0)$ instead of  bare intensity correlations $\left< I(t)I(0)\right>$ promises several advantages, as it automatically removes, \textit{e.g.}, a constant background intensity or stuck particles.

In Fig. \ref{fig:1}(c) we show the time-dependent mean square change of intensity (Eq. \eqref{eq:msi}) for different observation box sizes $L$. Although the system is quite simple, we find that generally, the intensity fluctuations undergo a variety of regimes, in time and with increasing box sizes, with well-defined features. Phenomenologically, $\left< \Delta I(t)^2\right>$ initially increases with time due to fluctuations in experimental data and particles entering or leaving the observation box. 
For long times, $\left< \Delta I(t)^2\right>$ eventually plateaus. In this regime, the  intensity $I(t)$ is fully uncorrelated with the initial intensity $I(0)$ and thus $\left< I(t)I(0)\right> \to \left< I(t)\right>\left<I(0)\right>$. Assuming that the system is stationary, and that the mean intensity within an observation box stays constant over time, the plateau value becomes twice the intensity variance, 
\begin{align}
\lim\limits_{t \to \infty} \left< \Delta I(t)^2\right> = 2 \left(\big< I^2\big> - \big< I \big>^2\right). \label{eq:msi_infty}
\end{align}
Calculating the intensity variance and plotting twice its value in dashed lines in Fig.~\ref{fig:1}(c) shows that it indeed lines up with the value of the plateau of the intensity fluctuations. 
We observe that intensity fluctuations are larger for larger box sizes $L$. The mean intensity level should scale with the mean number of particles, which increases in larger box sizes, explaining this effect. The number of particles and hence also its fluctuations are closely related to the intensity, as is expected and visualized in Fig. \ref{fig:1}(b).

\section*{A simple theory enables to quantify diffusion coefficients.}
To quantitatively rationalize the observed behaviour we develop a comprehensive theory. The intensity is related to the particle density distribution $\hat{\rho} = \sum_i \delta(\textbf{r}-\textbf{r}_i(t))$ with a convolution 
\begin{align}
I(\textbf{x},t) = \int d\textbf{x}' \text{PSF}(\textbf{x}-\textbf{x}') \hat{\rho}(\textbf{x}',t), \label{eq:I_PSF}
\end{align}
For simplicity, we neglect background intensity noise in this derivation. Note that a uniform background intensity level does not modify the derivation, and hence we take it to be zero.
We have introduced the particle spread function $\text{PSF}(\textbf{x})$, including both effects of the particle shape -- often called form factor -- and the point spread function of the microscope. \sophiem{missing some kind of ref here for PSFs etc, any idea sophie?}
In further steps, we will consider that the PSF is an isotropic function. 

To only consider contributions in a specific observation box, the intensity $I(\textbf{x},t)$ is given by
\begin{align}
I(t) = \int d\textbf{x}' \phi(-\textbf{x}') I(\textbf{x}',t). \label{eq:I_phi}
\end{align}
The box indicator function  $\phi(\textbf{x})$ here
describes the area from which the intensity is contributing, as we are interested in the \textit{total} intensity within the observation box. We first focus on intensities from a square subsection of the microscope image, indicated by the blue square in Fig. \ref{fig:1}(a).
In this case, intensity from all pixels within the box contribute, while pixels outside are not included. Therefore, $\phi(\textbf{x}) = \Theta(L/2-\left| x \right|) \Theta(L/2-\left| y \right|)$, where $\Theta(x)$ denotes the Heaviside step function, $\textbf{x}=(x,y)$ and $\left| \cdot \right|$ is the absolute value.
Here, the center of the observation box is chosen at the origin. In a homogeneous system as in ours, we average over different boxes on an image and so the position of the box center is not relevant. However, in inhomogeneous systems one could in principle perform region specific analysis~\cite{scipioni2018local}. 

Due to the convolutions in Eqs. \eqref{eq:I_PSF} and \eqref{eq:I_phi} it is advantageous to perform further steps in Fourier space. We use the convention $f(\textbf{k},t) =\int d \textbf{x} e^{-i \textbf{k}\cdot \textbf{x}} f(\textbf{x},t)$ for the Fourier transform and $f(\textbf{x},t) = (2 \pi)^{-d} \int d \textbf{k} e^{i \textbf{k}\cdot \textbf{x}} f(\textbf{k},t)$ for the inverse transform, where $d$ is the system's dimension. Note that for notation simplicity we explicitly specify at all instances if a function is taken in Fourier or real space through its dependence in $\textbf{k}$ or $\textbf{x}$ respectively. 
Combining Eqs. \eqref{eq:I_PSF} and \eqref{eq:I_phi}, we can express the intensity correlation as
\begin{align}
    \left< I(t) I(0) \right> = \int\!\!\frac{d\textbf{k}}{(2\pi)^d} \left|P_\text{eff}(\textbf{k})\right|^2  \big< {\hatrho}(\textbf{k},t) {\hatrho}(-\textbf{k},0) \big>, \label{eq:II}
\end{align}
where we introduced an effective convolution factor, $P_\text{eff}(\textbf{k}) = \text{PSF}(\textbf{k})\phi(\textbf{k})$, consisting of the Fourier transformed PSF and the box indicator function. Intensity correlations are thus closely related to the intermediate scattering function $f(\textbf{k},t) = \left< \hatrho(\textbf{k},t)\hatrho(-\textbf{k},t)\right>/N$. Equation~\eqref{eq:II} is a general expression describing the intensity correlations in \textit{a priori} any box shape or with any PSF, so any kind of visual rendering of a particle on the image.

To apply the general expression Eq.~\eqref{eq:II} to our experimental model system, we need to specify the PSF and the box indicator function. For the PSF, we consider a cross-section of an exemplary particle (see Fig. \ref{fig:1}(d)). As typical for confocal microscopy, the fluorescent intensity of a particle appears Gaussian. The experimental intensity data points and therefore the particle spread function can be approximated by a Gaussian function (red line), given by $\text{PSF}(\textbf{x}) = I_0 e^{-\textbf{x}^2/2\sigma^2}/(2 \pi \sigma^2)^{d/2}$ which is in Fourier space $\text{PSF}(\textbf{k}) = I_0 e^{\sigma^2 \textbf{k}^2/2}$ with the average intensity of one particle $I_0$ and the effective particle size $\sigma$ on the image. Note that $\sigma$ may be different than the actual particle size due to the point spread function of the microscope. 
The box indicator function is chosen as
a two-dimensional square box with side-length $L$ -- blue square in Fig. \ref{fig:1}(a), giving $\phi(\textbf{k}) = \frac{\sin(L k_x/2)}{k_x/2} \frac{\sin(L k_y/2)}{k_y/2}$, with $k_x$ and $k_y$ the Cartesian components of the wave vector $\textbf{k}$. Further particle spread functions and box indicator functions are discussed in Methods, Sec. C.

For the density-density correlation we use the exact expression \begin{align*}
    \left<\hatrho(\textbf{k},t) \hatrho(\textbf{k}',t') \right> =& \left(2 \pi\right)^{2d} \frac{N(N-1)}{V^2} \delta(\textbf{k}) \delta(\textbf{k}') \\&+ \left(2 \pi\right)^d \frac{N}{V} \delta(\textbf{k}+\textbf{k}') e^{-Dk^2 |t-t'|}
\end{align*} for $N$ purely diffusive, non-interacting particles moving in a $d$-dimensional volume $V$ \cite{velenich2008}. 
It is thus an appropriate and simple approximation in the model experiment with $d=2$ and a low particle density.

Combining these specifications we can determine the mean square intensity change \eqref{eq:msi} using Eq. \eqref{eq:II}, which after some algebra, simplifies to
\begin{align}
\left< \Delta I(t)^2 \right> &= 2 \left< I \right> I_0 \! \left[ f\!\left(\frac{4 \sigma^2}{L^2}\right)^{\!\!d} \!\!- f\!\left(\frac{4 D t+ 4\sigma^2}{L^2}\right)^{\!\!d}\right]\! \label{eq:I_Gaussian_Square} \\
\text{where } f(\tau) &= \sqrt{\frac{\tau}{\pi}} \left(e^{-1/\tau}-1 \right) + \text{erf}(\sqrt{1/\tau}),
\end{align}
and the mean intensity in the observation box is $\left< I \right> = I_0 N/V$. Further details are reported in the methods. Since $f(0) =1$, then when we consider point like particles ($\sigma = 0$) we recover the result for particle number fluctuations in a box~\cite{mackay2024} -- which we call the ``number countoscope'' limit. 

To align theory Eq.~\eqref{eq:I_Gaussian_Square} with experimental data, four parameters still need to be determined, $\left< I \right>$, $I_0$, $\sigma$, and $D$.
The mean intensity $\left< I \right>$ in the observation box can be straightforwardly obtained from the microscopy images, where the average is taken over all boxes and time.
The remaining three parameters are obtained through a fitting procedure. To do so, we first explore the long-time limit, \textit{i.e.}, the plateau values, as it should be agnostic of the dynamics of the system. 
In this limit, the theoretical result Eq.~\eqref{eq:I_Gaussian_Square} provides an expression for the plateau value, or more simply for the intensity variance as
\begin{align}
\left< I^2 \right> - \left< I \right>^2 = \left< I \right> I_0  f\left(\frac{4 \sigma^2}{L^2}\right)^{d}, 
\label{eq:msi_infty_theo} 
\end{align}
and is indeed independent of the diffusion coefficient $D$. We can use Eq.~\eqref{eq:msi_infty_theo} to fit
the intensity variance $\text{var}(I)$. 
Using a least squares fitting this determines the mean particle intensity as $I_0=2170$ and the mean particle width on the image as $\sigma = 1.05~\unit{\micro\metre}$. 
The results of the rescaled variance are shown for the experiments (blue dots) and a theoretical fit (gray line) in Fig. \ref{fig:1}(e). 
Having determined $\sigma$ and $I_0$ from the plateau values, Eq. \eqref{eq:I_Gaussian_Square} can be used to model the intensity fluctuations $\left< \Delta I(t)^2\right>$ and we obtain the self diffusion coefficient via a least-squares fit as $D = 0.076~\unit{\micro\metre^2/s}$. \sophieh{Error?!} 

Inserting all four parameters into Eq.~\eqref{eq:I_Gaussian_Square} (solid lines) gives the theoretical prediction as shown in Fig. \ref{fig:1}(c). Recall that the full set of seven curves shares the same three fitting parameters. The theoretical prediction accurately describes the experimental data (points) over all times and is the most accurate for small boxes. There are some deviations for larger boxes. These could be attributed to the lower statistical accuracy inherent to large boxes, since one can pave fewer large boxes than small ones on an image. They could also be attributed to the presence of collective effects  or experimental noise, which are currently not considered within the theory. 

The determined parameters agree well with our expectations. Most importantly, the value of the diffusion coefficient
is in agreement with the one obtained from the mean square displacement (MSD) calculated by tracking the particles, $D = 0.078~\unit{\micro\metre^2/s}$. Notice that compared to the Stokes-Einstein formula giving $D=0.107~\unit{\micro\metre/\second^2}$ for particle diffusion in the bulk, this value is slightly lowered due to increased hydrodynamic friction with the bottom wall~\cite{goldman1967slow}.
Further parameters also agree well with experimental reality. Taking the particle size as given by the location where the intensity Gaussian is at least $91\%$ of the maximum particle intensity, we can estimate the actual particle radius is roughly $a \simeq 2\sigma$. Since $a=2~\unit{\micro\metre}$ we find $\sigma  \simeq 1~\unit{\micro\metre}$ which is in good agreement with the mean particle width determined with the fit. 
This agreement is additionally verified by comparing the intensity distribution for an exemplary particle (blue points) with the Gaussian intensity profile (grey line), depending on the fitted parameters $I_0$ and $\sigma$, in Fig. \ref{fig:1}(d). Both intensity distributions are in reasonable agreement.

\section*{Theory highlights the presence of several regimes of intensity sensing.}


\begin{figure*}[tbhp]
    \includegraphics[width=.80\textwidth]{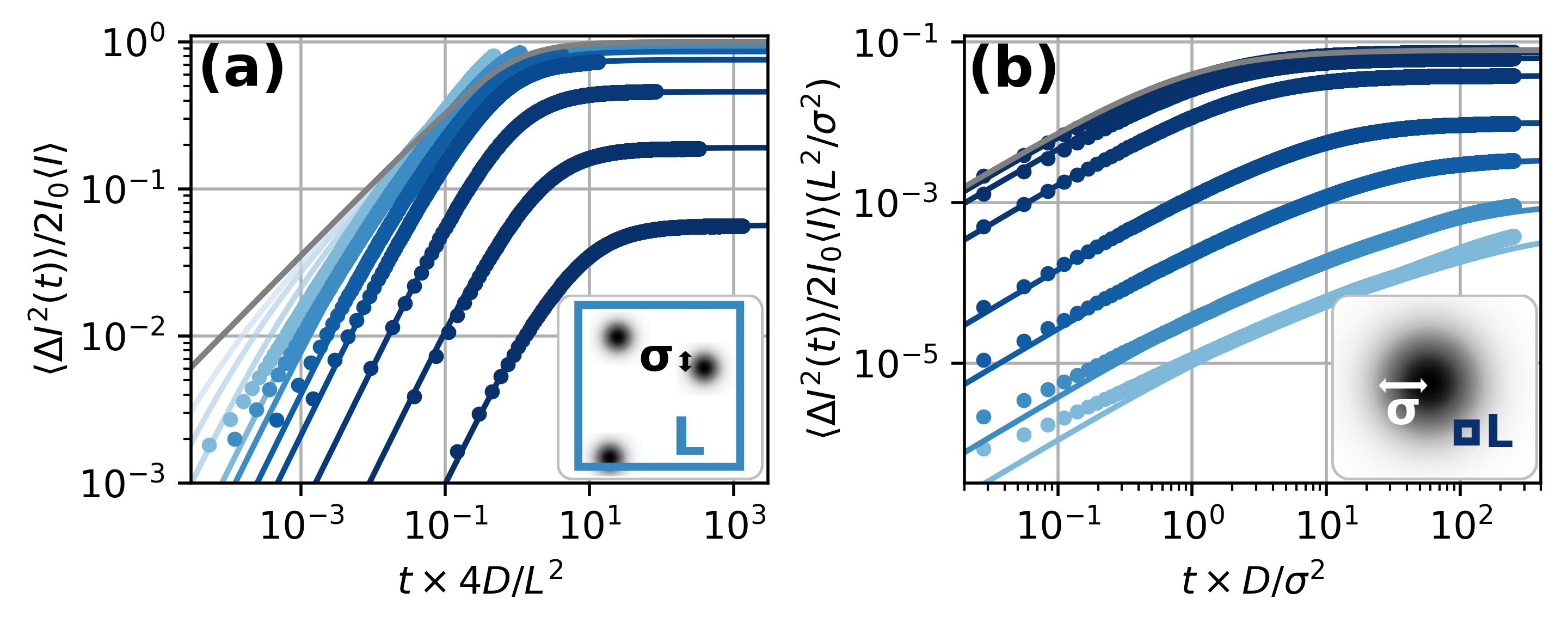}
    \caption{ \textbf{Different rescalings of intensity fluctuations $\left< \Delta I(t)^2\right>$ reveal different physical phenomena}.  (a) Case $\sigma \ll L $, where $\left< \Delta I(t)^2\right>$ is rescaled by the average intensity in a box $\left<I\right>$ and in time by the diffusive timescale $L^2/4D$. (b) Case $\sigma\gg L$, where $\left< \Delta I(t)^2\right>$ is normalized by the mean box intensity $\left<I\right>$ and the fraction of the particle the box covers $L^2/\sigma^2$, and in time by the diffusive timescale $\sigma^2/D$.
    Solid gray lines show limiting functions in the respective regimes. 
    The colour scale is the same as in Fig. \ref{fig:1}(c), with additional theory predictions in (a) for $L =  [128, 256, 512]~\unit{\micro\metre}$. The insets in (a,b) illustrate the ratio of relevant length scales, particle width $\sigma$ and box size $L$, for each regime.
    \sophieh{Note: the length ratio in the sketches are arbitrary not to scale (compared to the colour code).} 
    \sophiem{improve y label, missing t and $\sqrt{t}$ slopes}
    }
    \label{fig:rescaling}
\end{figure*}

Even at the level of a seemingly simple suspension of dilute diffusing particles, the intensity correlations both dynamic in Fig.~\ref{fig:1}(c) and static in Fig.~\ref{fig:1}(e) exhibit a rich phenomenology with different scaling laws at different time and lengthscales. 
Based on the theory developed, we can find characteristic timescales and intensity scales to reveal and understand these scaling laws further. To identify typical regimes of interest, it is clear from Fig.~\ref{fig:1}(e) that phenomena are quite different in the cases of small, respectively large, particles compared to box size $\sigma \ll L$, respectively $\sigma \gg L$. 


We start by the limit $\sigma \ll L$, where the particle width is much smaller than the observation box (see Fig. \ref{fig:rescaling}(a) for a sketch). First, we investigate static correlation functions. In this limit, we know from Eq.~\eqref{eq:msi_infty_theo}, that the plateau value scales as $\left< \Delta I(t)^2\right> \simeq 2\text{var}(I) = 2I_0 \left< I \right>$. 
Rescaling the variance $\text{var}(I)$ accordingly by $\left< I \right>$ as done in Fig. \ref{fig:1}(e) agrees with this prediction since the rescaled variance of the intensity becomes constant for $\sigma \ll L$. In this limit, particles correspond to point like particles, the $\text{PSF}$ becomes a delta distribution. Replacing $\text{PSF}(\textbf{x}) = I_0 \delta(\textbf{x})$ in Eq.~\eqref{eq:I_PSF} determines $I(\textbf{x},t) = I_0 \hat{\rho}(\textbf{x},t)$ and using Eq.~\eqref{eq:I_phi} we find the intensity is proportional to the number of particles, $I(t) = I_0 N(t)$, where $N(t)$ denotes the number of particles in the corresponding virtual observation volume. Since for a statistically uncorrelated suspension of particles $\text{var}(N) = \langle N \rangle$, then we expect that $\text{var}(I) \propto \langle I \rangle$ as well.  

Second, we investigate dynamic fluctuations of $I(t)$. 
The time required to reach the plateau of $\left< \Delta I(t)^2\right>$ is determined by the characteristic time to exchange particles between inside and outside the box. 
Since in this regime particle width is negligible compared to the observation box scale, this time is simply the time for the particle centres to move in and out of the box, $L^2/4 D$. Rescaling intensity fluctuations and time accordingly, we obtain a collapse onto a master curve for large enough observation boxes in Fig. \ref{fig:rescaling}(a). We show as a gray line the limiting ``number countoscope'' corresponding to infinitely small particles, $\sigma = 0$. The agreement collapse between numbers and intensities is mostly satisfied on large enough boxes, at long enough times, since at short times intensity fluctuations are dominated by finite-particle size effects -- which we will come to in the next paragraphs. At these long enough times, there emerges a $\sqrt{t}$-scaling which extends over all timescales when $\sigma = 0$. 
The observed $\sqrt{t}$-scaling results from diffusive particle motion across the boundaries of the observation box.

We next consider the limit $\sigma \gg L$, for which the observation box is much smaller than the particle width as sketched in Fig. \ref{fig:rescaling}(b). Again we first investigate static correlations. 
In this regime, the variance \eqref{eq:msi_infty_theo} becomes $\text{var}(I) = I_0 \left< I \right> \left(L/2\sqrt{ \pi} \sigma\right)^{d}$. The first factor $I_0 \left< I\right>$ recovers the scaling observed in the previous limit $\sigma \ll L$ and the second factor accounts for the reduced intensity fluctuations due to the small box size compared to the particle width. An heuristic approximation of this factor is the fraction of the particle that is covered by the box, $V_\text{box}/V_\text{particle}$, recovering the expected scaling $\propto L^d/\sigma^d$ of the variance.

We also investigate dynamic correlations in this limiting regime $\sigma \gg L$. 
The time required for $\langle \Delta I^2(t) \rangle$ to reach the plateau is again determined by diffusive particle motion, with the relevant length scale being set by the particle width $\sigma$. This defines a characteristic time scale as $D/\sigma^2$.
By rescaling the time and intensity fluctuations accordingly, we observe a collapse onto the limiting function as $\sigma/L \to \infty$ (gray line in Fig. \ref{fig:rescaling}(b)). Approaching the limit is, in practice, primarily constrained by the spatial resolution of the microscope. The smallest accessible box size corresponds to a single pixel. 
For short times, we predict from Eq. \eqref{eq:I_Gaussian_Square} and find a linear increase of the rescaled mean square change in intensity as $d D t/2 \sigma^2$. Remarkably, in this regime the mean squared intensity differences scale like the mean squared displacements, 
\begin{equation}
    \frac{\langle \Delta I^2(t)\rangle }{2I_0\langle I \rangle } \frac{4\pi\sigma^2}{L^2} = \frac{\langle \Delta \textbf{x}^2(t) \rangle}{4 \sigma^2}.
\end{equation}
This originates from the fact that at these small timescales, we can essentially think of the box as ``diffusing'' on the particles' PSF, yielding this peculiar scaling law. 
The short time regime persists across box sizes albeit with slightly different scaling laws, as $d\pi^{-1/2} \flatfrac{Dt}{\sigma L}$ for $\sigma \ll L$, demonstrating this behaviour is quite universal for short time scales. 
 
\sophieh{What about including the transition times (1 and 2)? Potentially indicate them in Figures?} \sophiem{Maybe you can add just a paragraph explaining how we go from one to another regime?}

\begin{figure*}[htbp]
    \centering
    \includegraphics[width=1.\textwidth]{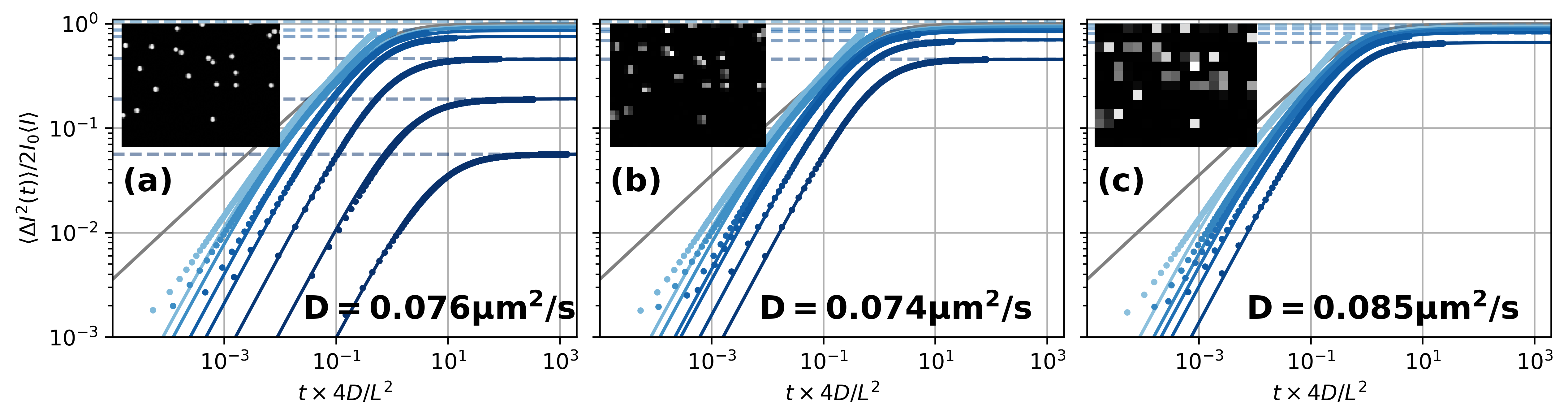}
    \caption{\sophieh{Add scale bar (and potentially L-box) to the insets.} \textbf{Intensity fluctuations capture diffusion coefficients at different image resolutions.} $\langle \Delta I(t)^2\rangle$ for different resolutions of images: \textbf{(a)} 512$\times$512 pixel (full resolution, $l_\text{px} = 0.91~\unit{\micro\metre}$), \textbf{(b)} 128$\times$128 pixel ($1/16$ of the full resolution, $l_\text{px} = 3.6~\unit{\micro\metre}$), and \textbf{(c)} 64$\times$64 pixel (1/64 of the full resolution, $l_\text{px} = 7.28~\unit{\micro\metre}$). The data is rescaled as in Fig.~\ref{fig:rescaling}(a). The colour scale is the same as in Fig. \ref{fig:1}(c). The insets show a section of the optical image with the corresponding resolution. The diffusion coefficients are determined via fitting.
    }
    \label{fig:quality}
\end{figure*}

\section*{The method is robust in the low detection limit}

Since intensity counting is based on comparing intensity fluctuations in boxes and is essentially agnostic of individual particles, the method should in principle be applicable to systems where individual particles are not resolved -- much like FCS, ICS and DDM~\cite{cerbino2008differential}. 
To test the robustness of the approach, we artificially decrease the resolution of the microscope images.
To do so, starting from (a) our full $1\times1$ pixel resolution, we divide the image into non-overlapping squares of (b) $4\times4$, (c) $8\times8$ pixels and assign the mean intensity of each square to the value of a ``single pixel'' of these downsampled image. This procedure has two effects: It increases the length of a single pixel $l_\text{px}$ from (a) $0.91~\unit{\micro\metre}$ to (b) $3.64~\unit{\micro\metre}$, and (c) $7.28~\unit{\micro\metre}$ and decreases the image resolution from (a) $512\times512$ to (b) $128\times128$ and (c) $64\times64$ pixels. For a visual representation of the changes, we present representative images at each step of the downsizing process in the insets in Fig.~\ref{fig:quality}.

We then calculate the mean square change of the intensity for the downsampled images and present the results in Fig. \ref{fig:quality}. We choose to present the results with the rescaling that corresponds to the situation where $\sigma \ll L$, since in this downsizing process we typically do not resolve individual particles anymore. Note, that Fig.~\ref{fig:quality}(a) reproduces Fig.~\ref{fig:rescaling}(a) for pedagogic purposes. Interestingly, downsizing preserves the main aspects of the curves. The most notable difference is that naturally small observation boxes are not accessible due to the increased pixel size, \textit{i.e.}, there are fewer dark blue lines. On the largest boxes however, the features of the intensity fluctuations are quite similar regardless of the initial image quality (lighter blue colours). 

Using a similar fitting procedure as for Fig.~\ref{fig:1}(c), we also determine the diffusion coefficients as, from (a) $0.076~\unit{\micro\metre^2/\second}$, (b) $0.074~\unit{\micro\metre^2/\second}$, and (c) $0.085~\unit{\micro\metre^2/\second}$.
For a small downsampling in (b) the diffusion coefficients obtained agree, but there is some deviation at the $8\times$ downsampling level. One reason is that the procedure for fitting the variance (see Fig. \ref{fig:1}(e)) becomes more challenging as the decreasing resolution reduces the number of data points in the $\sigma \gg L$ regime. Consequently, the uncertainty in the particle width increases as the variance becomes independent of $\sigma$ in the limit $\sigma \ll L$. In addition, this restricts the fitting to large boxes which are more sensitive to statistical noise. 
Note that exploiting prior knowledge of $\sigma$ -- essentially, keeping the same value of $\sigma$ for all three resolutions, reduces the deviations. In this case, we determine the diffusion coefficients as, from (b) $0.076~\unit{\micro\metre^2/\second}$ and (c) $0.075~\unit{\micro\metre^2/\second}$. In a real experiment where resolution is low as in (c), one may not have a priori access to $\sigma$ and thus alternative fitting procedures may have to be found.
Overall, this shows that intensity counting remains robust, enabling determination of the diffusion coefficient, even when segmentation of individual particles is not possible. Hence, these results are promising for applications to a broader set of experimental conditions.


\section*{Discussion and conclusion}

In this work, we have analysed intensity fluctuations within virtual observation boxes in real space to quantify particle dynamics. We derived a theoretical description of the mean square change of the intensity, $\langle \Delta I^2(t) \rangle$, and validated it in an experimental model system. Using this approach, we determine the diffusion coefficient $D$, which shows excellent agreement with results obtained from the MSD of the tacked particles. Our analysis reveals distinct temporal regimes that depend on the ratio of the particle width to the virtual box size: at short timescales, $\langle \Delta I^2(t) \rangle$ scales with the MSD, while at longer timescales, it scales with its square root.
Much like FCS, ICS, and DDM, our method is robust and does not require resolving individual particles, making it widely applicable. Our real-space approach is computationally cheap, since it avoids Fourier transforms. In addition, working in real-space allows one to relate to physical phenomena in an easier way.

While our investigation was centred around the 2D experimental model system imaged with a confocal microscope, the introduced theoretical framework is quite general. By essence, it adjusts the shape of the box $\phi(\textbf{x})$ and particle $\text{PSF}(\textbf{x})$ at will. We provide an overview of common shapes in Tab. \ref{tab:particle}, based on typical shapes that one might need to describe various experimental systems (see also Methods, Sec. C). With this framework, we can for example recover some theories for FCS, where the box has a Gaussian shape corresponding to the Gaussian illuminated region, and the particle is considered point like~\cite{elson1974fluorescence,hofling2011fluorescence}. 
We hence anticipate that our method can be adopted for other samples or microscopy techniques, such as total internal reflection fluorescence (TIRF) and bright-field microscopy. Furthermore, given we operate in real space, our approach could in principle be easily extended to probe large scale collective effects -- circumventing some challenges such as edge-effects~\cite{carter2025measuring}. \sophiem{add more refs here} 

\sophieh{outlook on more practical details (e.g. 3D, drift, high densities, etc)}

\section*{Methods}
\subsection{Experiments}
\sophiem{remember to figure out what the cell is made of}
Experiments are carried out with sulfate-modified fluorescent polystyrene particles with 
$4~\unit{\micro\metre}$ nominal diameter (Invitrogen, F8859). Particles are dispersed in ultra-pure deionized water (18.2 $\mathrm{M\Omega \cdot cm^{-1}}$) at a concentration of roughly $0.002 \text{wt} \%$. Fluorescent particles are imaged using a Leica SP8 Confocal Laser Scanning Microscope (CLSM) with a 20$\times$ objective (0.75 NA), while the pinhole is set to 1 Airy unit. The resulting field of view is $465.62 \times 465.62 ~\unit{\micro\metre}^2$. 
The particles are excited at 514 nm and their emission is collected at wavelengths between 525 and 600 nm. Optical properties are selected to avoid saturation of the detector. A resonant scanner (8 kHz) is used at an acquisition rate of 2.5 
frames per second at a resolution of 512 × 512 pixels. To accurately resolve the plateaus of the intensity fluctuations (see Fig \ref{fig:rescaling}) videos are recorded for 
$5~\unit{\hour}$. 
To minimize changes in average intensity over the course of an experiment, the autofocus feature of the Leica LAS X software is used every 600 frames.



\subsection{Data analysis}

Data analysis is performed directly on the microscopy videos. The field of view, assumed to be fixed in time, is divided into $M$ square tiles based on the specified observation box length $L$ and a specified separation width $\delta L$ between the boxes. For simplicity, we only consider box lengths, separations, and positions that align with pixels.
We then sum the intensity of each pixel that falls within each observation box, generating a time series of intensities $I_i(t)$ for the box indices $i=1,...,M$. Statistical quantities such as $\left< \Delta I(t)^2 \right>$ are estimated by first averaging over time origins $t_0$ and then using arithmetic means over all of the boxes $ \big< \left< \Delta I(t)^2 \right>_{t_0} \big>_{i}$. The average particle intensity $\left< I \right>$ and variance $\left< I^2 \right> -\left< I \right>^2$ are estimated using the sample mean and the unbiased sample variance, determined from all boxes and frames.

\subsection{Theory}

For pedagogy purposes, we present the detailed theoretical framework here in a self-contained manner. 
The intensity $I(\textbf{x},t)$ of a system is given by the sum of the intensity distribution of each individual particle, $I(\textbf{x},t) = \sum_i \text{PSF}(\textbf{x} - \textbf{x}_i(t))$. Here we assume that every particle has the same intensity profile, the particle spread function $\text{PSF}(x)$ which includes the effects of the optical imaging system (usually included in the point spread function) and of the particle shape. Mathematically, the intensity can hence be expressed as
a convolution of the $\text{PSF}(\textbf{x})$ and the density distribution $\hat{\rho} = \sum_i \delta(\textbf{x}-\textbf{x}')$, see Eq. \eqref{eq:I_PSF}.
We are interested in the intensity fluctuations within a given area, the so-called observation box, which is given by an integral over the corresponding volume $\int_{V_\text{box}} \!\!d\textbf{x} I(\textbf{x},t)$ or more generally by a convolution
\begin{align}
    I(\tilde{\textbf{x}},t) = \int d\textbf{x} \; \phi(\tilde{\textbf{x}}-\textbf{x}) I(\textbf{x},t) \equiv (\phi * I(t))(\tilde{\textbf{x}}), \label{eq:I(t)}
\end{align}
where $\phi(\textbf{x})$ is the indicator function of the observation box with volume $V_\text{box} = \int d \textbf{x} \phi(\textbf{x})$. In the simple case of a one dimensional square box of length $L$, Eq. \eqref{eq:I(t)} becomes $$I(\tilde{x},t) = \int_{\tilde{x}-L/2}^{\tilde{x}+L/2} dx \; I(x,t)$$ and hence $\phi(x) = \Theta(\frac{L}{2} - \left| x \right|)$, where $\Theta(x)$ denotes the Heaviside step function.

The combination of both convolutions Eq. \eqref{eq:I_PSF} and Eq. \eqref{eq:I(t)} yields
\begin{align}
    I(\tilde{\textbf{x}},t) =  &\int d\textbf{x} \phi(\tilde{\textbf{x}}-\textbf{x}) \int d\textbf{x}' \text{PSF}(\textbf{x} - \textbf{x}') \hat{\rho}(\textbf{x}',t) \label{eq:NR1} \\
    = &\int \frac{d \textbf{k}}{(2 \pi)^d} \hatrho(\textbf{k},t) e^{ i \textbf{k}\cdot \tilde{\textbf{x}}} \int d\textbf{x} \phi(\tilde{\textbf{x}}-\textbf{x}) e^{- i \textbf{k}\cdot (\tilde{\textbf{x}}-\textbf{x})} \nonumber \\
    &\times \int d\textbf{x}' \text{PSF}(\textbf{x} - \textbf{x}') e^{- i \textbf{k}\cdot (\textbf{x}-\textbf{x}')} \\
    = &\int \frac{d \textbf{k}}{(2 \pi)^d} e^{ i \textbf{k}\cdot \tilde{\textbf{x}}}  \hatrho(\textbf{k},t) \phi(\textbf{k}) \text{PSF}(\textbf{k}), \label{eq:NR2}
\end{align}
where $d$ denotes the spatial dimension and we replaced the density distribution in Eq. \eqref{eq:NR1} by its expression in Fourier space, $\hat\rho(\textbf{x},t) = \int d\textbf{k} e^{i \textbf{k} \cdot \textbf{x}} \hatrho(\textbf{k},t)/(2\pi)^d$.
Alternatively, Eq. \eqref{eq:NR2} follows directly after applying the convolution theorem $\mathcal{F}[g*h] = \mathcal{F}[g]\mathcal{F}[h]$ twice, where $\mathcal{F}$ indicates the Fourier transform.
Note that for systems with constant density distribution, the intensity is independent of the position of the observation box. So for simplicity, we assume $\tilde{\textbf{x}}=0$ in the following.

The temporal intensity correlation $\left< I(t) I(t') \right>$ of an observation box with itself is thus
\begin{align}
     &\int\!\!\frac{d\textbf{k}}{(2\pi)^d} P_\text{eff}(\textbf{k}) \int\!\!\frac{d\textbf{k}'}{(2\pi)^d} P_\text{eff}(\textbf{k}') \left< \hatrho(\textbf{k},t) \hatrho(\textbf{k}',t') \right>,
\end{align}
where we combined the effects of the particle intensity and of the observation box in one effective convolution factor $P_\text{eff}(\textbf{k}) = \phi(\textbf{k}) \text{PSF}(\textbf{k})$. 
Further exploiting spatial homogeneity of the system allows to simplify 
\begin{align}
     \left< I(t) I(t') \right>
     &=\int\!\!\frac{d\textbf{k}}{(2\pi)^d}  \left|P_\text{eff}(\textbf{k})\right|^2 \left< \hatrho(\textbf{k},t) \hatrho(-\textbf{k},t') \right>, \label{eq:NR3}
\end{align}
which is equal to Eq. \eqref{eq:II} from the main text for $t'=0$.

To proceed, we need to specify the density-density distribution function. In $k$-space the density is given as $\hatrho(\textbf{k},t) = \sum_j e^{i \textbf{k}\cdot \textbf{x}_j(t)}$ and the position of the $j$th particle $\textbf{x}_j(t)$ can be determined by the solution of the corresponding Equation of motion.

We here focus on purely diffusive, non-interacting particles that satisfy the Langevin equation 
\begin{align}
    \frac{d \textbf{x}_j}{d t} = \sqrt{2D} \boldsymbol{\eta}_j(t), \label{eq:Langevin}
\end{align} 
where $\boldsymbol{\eta}(t)$ is a Gaussian white noise with zero mean $\left< \boldsymbol{\eta}_i(t) \right> =0$, correlation $\left< \boldsymbol{\eta}_i(t) \boldsymbol{\eta}_j(t') \right> = \delta_{ij} \delta(t-t') \mathbb{1}$ and $\mathbb{1}$ denotes the unit matrix. 
Due to its simplicity, Eq. \eqref{eq:Langevin} can be solved analytically as $$\textbf{x}_j(t) = \textbf{x}_j(0) + \sqrt{2D} \int_0^t dt' \boldsymbol{\eta}_j(t')$$ and allows to express the density-density correlation function exactly (see e.g. Reference \cite{velenich2008}) as
\begin{align}
    \left< \hatrho(\textbf{k},t) \hatrho(\textbf{k}',t')\right> =&  \left(2 \pi\right)^{2d} \frac{N(N-1)}{V^2} \delta(\textbf{k}) \delta(\textbf{k}') \nonumber \\
    &+ \left(2 \pi\right)^d \frac{N}{V} \delta(\textbf{k}+\textbf{k}') e^{-Dk^2 |t-t'|}, 
\end{align}
where the brackets $\left< \cdot \right>$ indicate an average over initial positions $\textbf{x}_i(0)$ and over noise $\boldsymbol{\eta}(t)$.

The corresponding intensity fluctuations \eqref{eq:NR3} in a diffusive system are 
\begin{align}
    \left< I(t) I(0) \right> = &N(N-1)  I_0^2  \frac{V_\text{box}^2}{V^2}\nonumber \\
    &+ \frac{N}{V} \int \frac{d\textbf{k}}{(2\pi)^d} \left|P_\text{eff}(\textbf{k})\right|^2 e^{- D t \textbf{k}^2} \label{eq:<II>}
\end{align}
where we exploited $\text{PSF}(\textbf{k}\!=\!\!0) = \int d\textbf{x} \text{PSF}(\textbf{x}) = I_0$ is the mean particle intensity, and similarly $\phi(\textbf{k}\!=\!0) = V_\text{box}$.

In practice, it turns out to be useful to consider fluctuations of the intensity change, as introduced in Eq. \eqref{eq:msi} in the main text,
$ \left< \Delta I(t)^2\right> = \left< (I(t)-I(0))^2\right> =  2 \left< I^2 \right> -  2 \left< I(t) I(0) \right>$, where we assume stationarity such that the mean intensity is constant $\left< I \right> \equiv \left< I(0) \right>= \left< I(t) \right>$.
In case of the non-interacting purely diffusive system we determine, using Eq. \eqref{eq:<II>}, 
\begin{align}
     \frac{\left< \Delta I(t)^2\right>}{2 N/V} = \int \frac{d\textbf{k}}{(2\pi)^d} \left|P_\text{eff}(\textbf{k})\right|^2 \left(1 - e^{- D t k^2} \right).
\end{align}
At this stage, neither the particle intensity $\text{PSF}(\textbf{k})$ nor the shape of the observation box $\phi(\textbf{k})$ have been specified and can be chosen as applicable. We give reasonable choices for these functions in the following.

 \paragraph*{Particle spread functions}
The intensity distribution of a particle can take many different shapes. 
Recall that we chose the PSF from normalized functions, rescaled such that its integral $\int d\textbf{x} \text{PSF}(\textbf{x}) = I_0$ is the mean particle intensity. 
An overview of potential particle types, including their $\text{PSF}(\textbf{x})$, its inverse value at the particle centre $\textbf{x}=0$ and its Fourier transform $\text{PSF}(\textbf{k})$ in Tab. \ref{tab:particle}. The latter is required to determine the intensity correlation function, Eq. \eqref{eq:II}
We will now explicitly introduce several reasonable and mathematically feasible distributions. 

\begin{table*}[t]
    \centering
    \begin{tabular}{lc|cccc}
        \textbf{Particle  -} &\textbf{\!\!type}& & $\text{PSF}(\textbf{x})/I_0$& $I_0/\text{PSF}(\textbf{x}=0)$ & $\text{PSF}(\textbf{k})/I_0$ \\
        \hline
        &&&&\\[-0.5ex]
        Pointlike &  \raisebox{-0.15cm}{\includegraphics[width=0.04\textwidth]{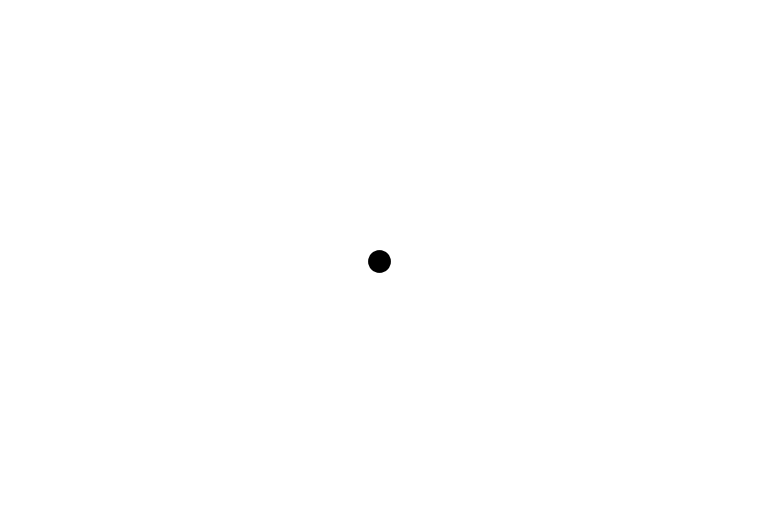}} &\raisebox{-0.15cm}{\includegraphics[width=0.04\textwidth]{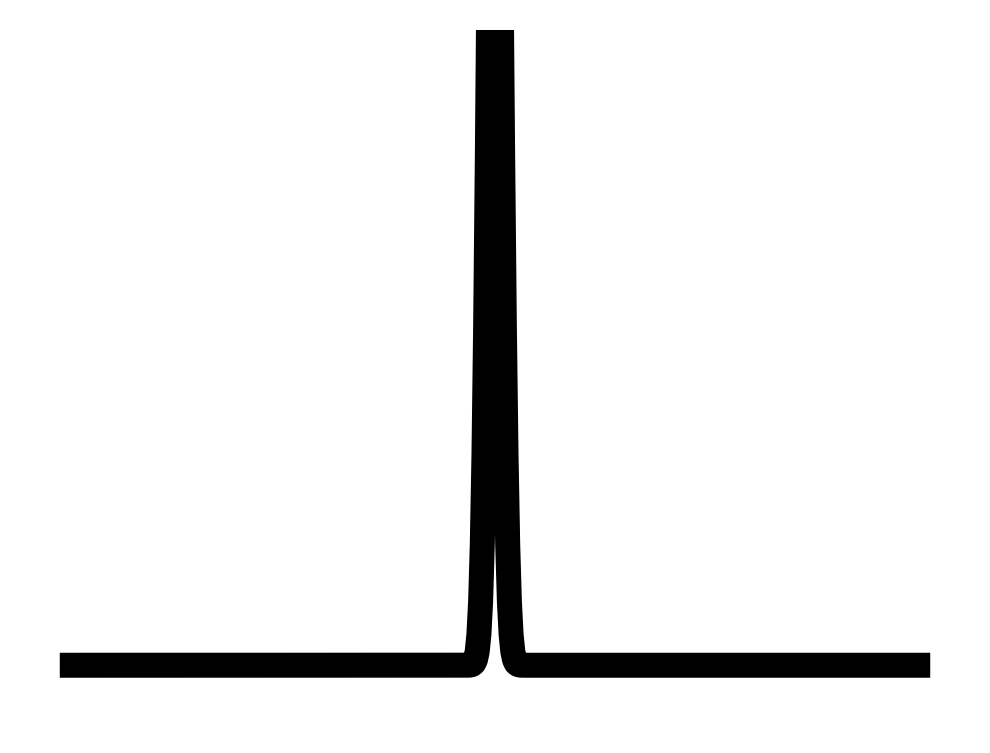}} & $ \delta(x_\alpha)$ & NaN & 1 \\[2ex]
        Gaussian & \raisebox{-0.15cm}{\includegraphics[width=0.04\textwidth]{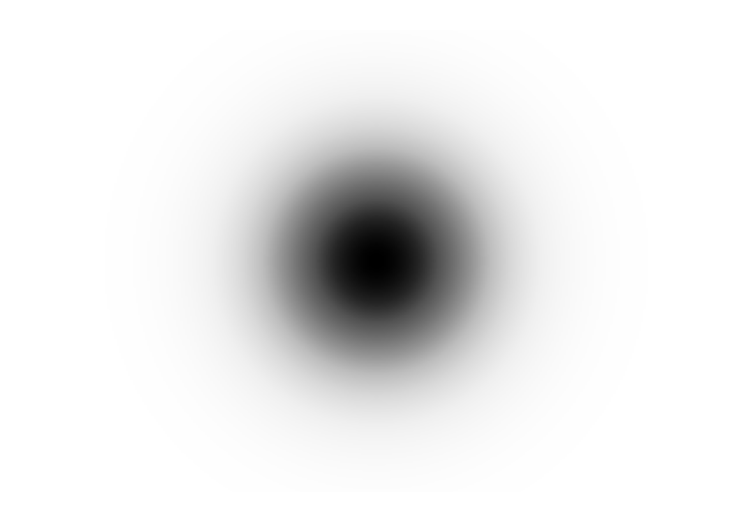}}& \raisebox{-0.15cm}{\includegraphics[width=0.04\textwidth]{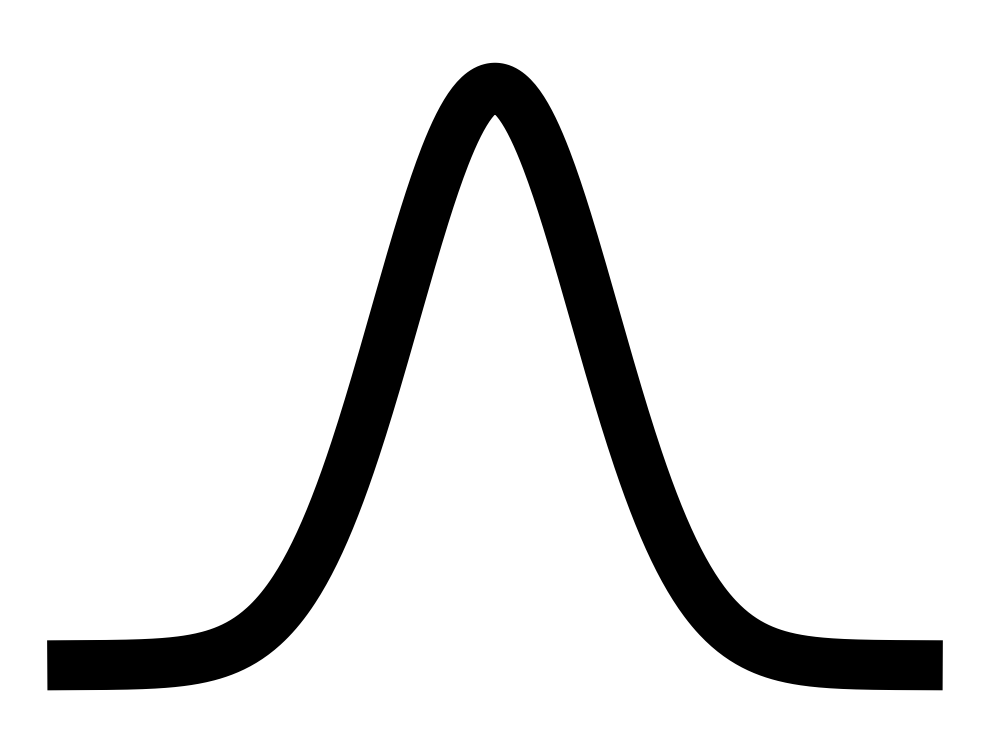}} & $ \frac{1}{\sqrt{2 \pi \sigma^2}} \; e^{-{x_\alpha}^2/(2\sigma^2)}$ & $ \sqrt{2 \pi \sigma^2}$ & $e^{-k_\alpha^2\sigma^2/2}$ \\[2ex]
        Square &  \raisebox{-0.15cm}{\includegraphics[width=0.04\textwidth]{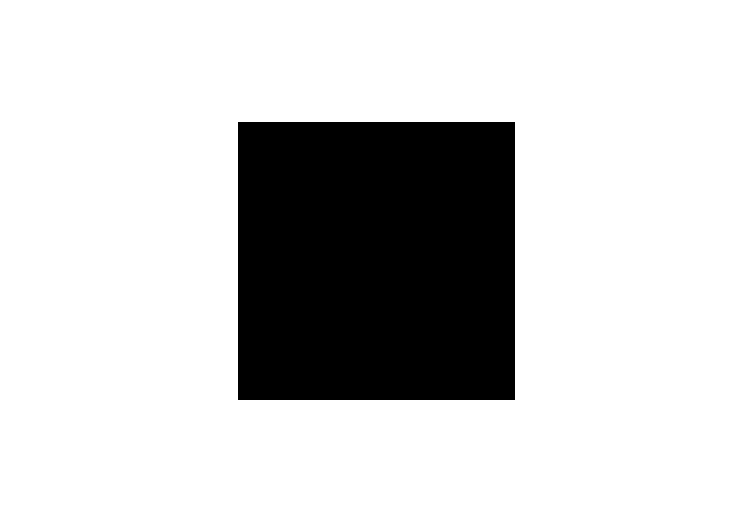}}&\raisebox{-0.15cm}{\includegraphics[width=0.04\textwidth]{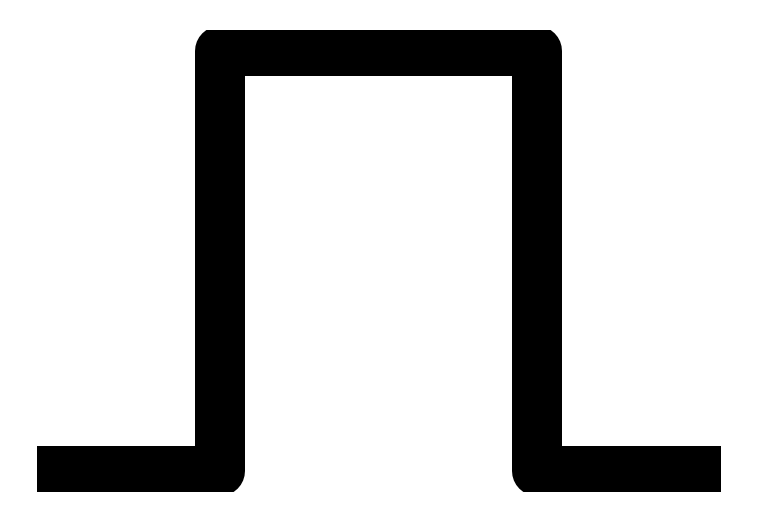}} & $\frac{1}{\sigma} \Theta(\frac{\sigma}{2} - |{x}_\alpha|)$ & $\sigma$ & $\frac{\sin(k_\alpha \sigma/2)}{k_\alpha \sigma/2}$ \\[2ex]
        Exponential & \raisebox{-0.15cm}{\includegraphics[width=0.04\textwidth]{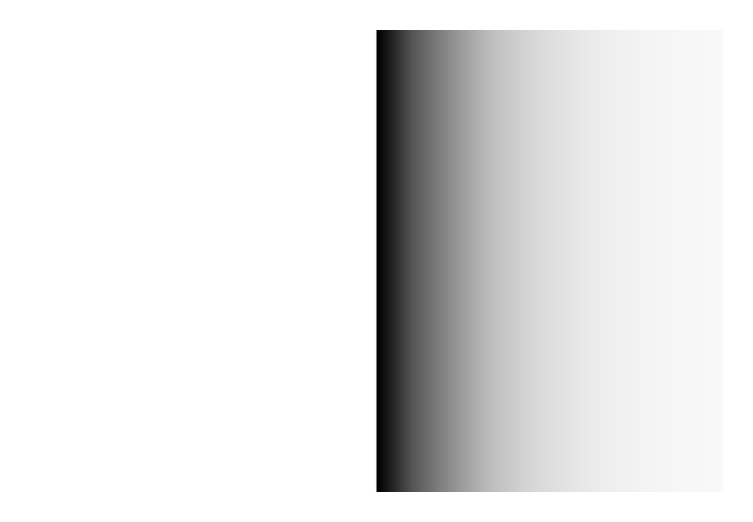}} & \raisebox{-0.15cm}{\includegraphics[width=0.04\textwidth]{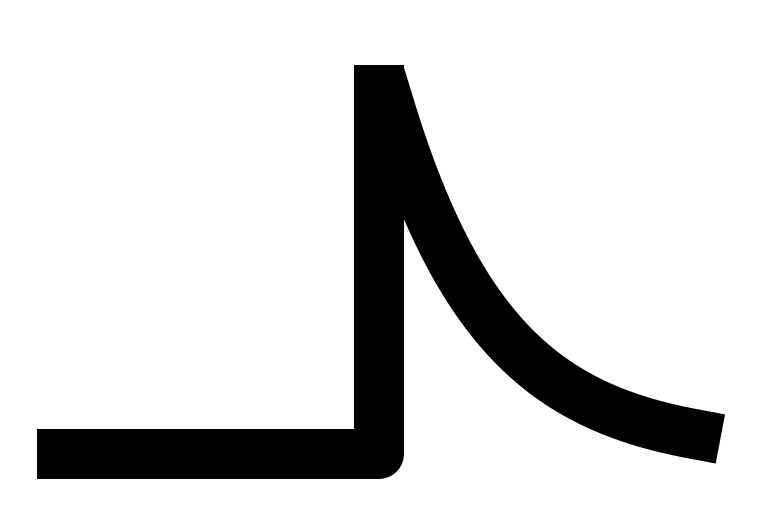}} & $\frac{1}{\sigma}\Theta(x_\alpha) e^{-x_\alpha/\sigma}$ & $\sigma$ & $(1+i k_\alpha \sigma )^{-1}$ \\[2ex]
        Circle (2D) & \raisebox{-0.15cm}{\includegraphics[width=0.04\textwidth]{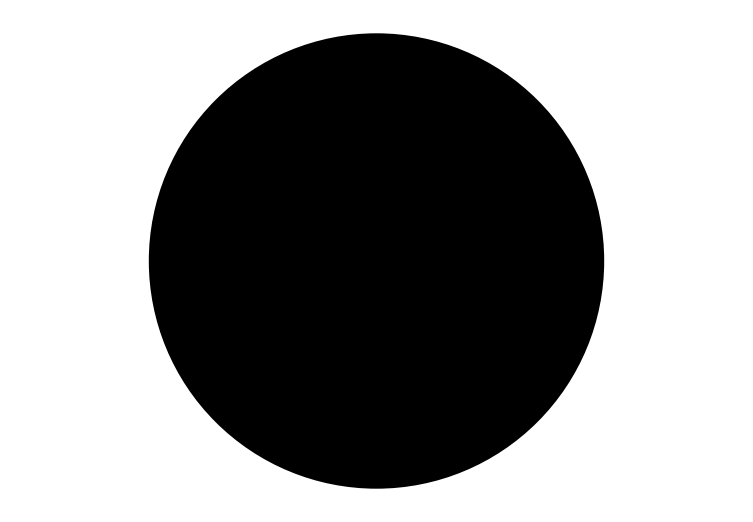}} & \raisebox{-0.15cm}{\includegraphics[width=0.04\textwidth]{Heaviside.png}} & $\frac{4}{\pi\sigma^2} \Theta(\frac{\sigma}{2} - |\textbf{x}|)$ & $\frac{\pi\sigma^2}{4}$ & $\frac{4}{k \sigma} J_1(k \sigma/2)$ \\[2ex]
        Sphere (3D) & \raisebox{-0.15cm}{\includegraphics[width=0.04\textwidth]{Circle.png}} & \raisebox{-0.15cm}{\includegraphics[width=0.04\textwidth]{Heaviside.png}} & $\frac{6}{\pi\sigma^3} \Theta(\frac{\sigma}{2} - |\textbf{x}|)$ & $\frac{\pi\sigma^3}{6}$ & $\frac{24}{k^3\sigma^3} \left( \sin\left(\frac{k \sigma}{2}\right) - \frac{k \sigma}{2} \cos\left(\frac{k \sigma}{2}\right) \right)$ 
    \end{tabular}
    \caption{\textbf{Particle spread function for different particle geometries.} For each particle type (illustrated) we show: the real-space $\text{PSF}(\textbf{x})$ (first column, including a sketch of the distribution), its inverse value at the particle center, $1/\text{PSF}(\textbf{x}=0)$ (second column), and the Fourier-space particle spread function $\text{PSF}(\textbf{k})$. Unless otherwise stated in the particle type, the particle spread functions are spatially separable and given in one dimension, with the index $\alpha$ indicating the corresponding Cartesian coordinate. In the expressions, $J_n$ denotes a Bessel function of the first kind and $\Theta$ is the Heaviside step function. While this table presents PSFs, the same functions can be directly adapted for use as observation box shapes $\phi$ by following the three-step procedure detailed in the main text.}
    \label{tab:particle}
\end{table*}

Very small particles in theory 
might be described as point-like by a Dirac delta distribution, $\text{PSF} = I_0 \delta(\textbf{x})$. In practice, however, particles are never truly point-like as a result of their finite size and the finite resolution of the optical system. Both effects yield a spatially extended intensity distribution. The delta-peak model is still relevant, as it probes the presence of particles and hence particle number fluctuations \cite{mackay2024}.

When the detected intensity distribution is dominated by the particle shape, \textit{e.g.,} for particles large compared to the point spread function of the imaging system, the observed intensity profile closely mirrors the geometric form factor of the particle. For spherical colloids, this yields a spherically symmetric intensity distribution $\text{PSF} = 4 I_0 \Theta(\flatfrac{\sigma}{2} - |\textbf{x}|)/\pi \sigma^2$, describing circles of diameter $\sigma$ as imaging usually implies a projection to the two dimensional plane.
Note that similar step-like intensity profiles can also arise artificially during post-processing, \textit{e.g.,} to enhance particle visibility and due to detector saturation, which truncates the intensity signal.

For particles much smaller than the diffraction limit, the detected intensity distribution is primarily determined by the point spread function. In such cases, the particle intensity can be approximated by a Gaussian profile, yielding $\text{PSF}(\textbf{x}) = I_0 (2 \pi \sigma^2)^{-d/2} e^{-\textbf{x}^2/2\sigma^2}$, where $d$ again represents the spatial dimension. This approximation is especially valid in confocal and fluorescence microscopy, which is why we adopted it for our model system imaged using a LSCM.
In general, optical setups, such as brightfield microscopy or TIRF, can produce more complex intensity profiles, including Airy disks. 
\sophieh{Figure out if there is a simple example to be included or we discard this to future work. So far no simple/analytical intensity correlation found.} 

Most PSFs are spatially separable, \textit{i.e.}, they can be expressed as a product of one-dimensional functions for each spatial coordinate. For this reason we list these one-dimensional contributions in Tab.~\ref{tab:particle}. Notable exceptions include circular and spherical intensities, which are inherently non-separable because of their rotational symmetry and are hence expressed explicitly.  

\paragraph*{Observation box shapes}
The observation box indicator functions $\phi(\textbf{x})$ are structurally quite similar to the particle spread functions discussed above. This allows us to reuse the one-dimensional distributions listed in Tab.~\ref{tab:particle} to describe their various shapes. The conversion from PSFs to $\phi$ can be carried out as follows: 1) Select the desired box shape from the particle types listed in Tab.~\ref{tab:particle}. 
2) Multiply the first by the second column in Tab.~\ref{tab:particle}. This rescales $\text{PSF}(\textbf{x})$ by its maximum intensity, typically located at the particle centre $x=0$, and thereby accounts for the different normalizations of PSF and $\phi$. The Fourier-space representation, $\phi(\textbf{k})$ is obtained similarly by multiplying the third by the second column. 
3) Replace the variable $\sigma$ with $L$, since the particle width and box size represent distinct relevant length scales.
This conversion generally applies, except for the point-like particle type, whose maximum intensity diverges. 

As an explicit example, we consider a Gaussian-shaped observation box, which could resemble the illuminated region in an FCS experiment. Following the above procedure, we determine $\phi(\textbf{k}) = (2 \pi L^2)^{d/2} \exp(-\textbf{k}^2 L^2/2)$.
For a two-dimensional square box, as used for the experimental model system in the main text, we obtain $\phi(\textbf{k}) = 4\flatfrac{\sin(L k_x/2) \sin(L k_y/2)}{k_x k_y}$.

\bibliography{biblio}

\end{document}